\documentclass{article}

\usepackage{auto-pst-pdf}
 
\usepackage{amsmath}
\usepackage{amsfonts}
\usepackage{booktabs}
\usepackage{graphicx}    

\newcommand{\zip}{\mathrm{ZIP}} 
\newcommand{\zinb}{\mathrm{ZINB}} 

\begin{document}
\title{Spline regression for zero-inflated models}
\author{T.~Opitz, P.~Tramini, N.~Molinari}
\maketitle
\begin{abstract}
We propose a regression model for count data when the classical GLM approach  is too rigid due to a high outcome of zero  counts and a nonlinear influence of  continuous covariates. 
Zero-Inflation is applied to take into account the presence of excess zeros with separate link functions for the zero and the nonzero component. Nonlinearity in covariates is captured by  spline functions based on B-splines. 
Our  algorithm relies on  maximum-likelihood estimation and allows for adaptive box-constrained knots, thus improving the goodness of the spline fit and allowing for detection of sensitivity changepoints. 
A simulation study substantiates the numerical stability of the algorithm to infer such models. The AIC  criterion is shown to  serve well for model selection,  in particular if non-linearities are weak such that BIC tends to overly simplistic models.   We fit the introduced models to real data of children's dental sanity, linking caries counts with the so-called Body-Mass-Index (BMI) and  other socio-economic factors. This reveals  a puzzling  non-monotonic influence of BMI on caries counts which is yet to be explained by clinical experts.  
\end{abstract}
\maketitle
\section{Introduction}
Classical log-linear regression for Poisson count data results in an inadequate fit to data if at least one of the two basic assumptions is violated: First, even conditionally on the covariates,  the counted events may not arise  independently and with identical distribution.  In real data, this typically leads to overdispersion, breaking the mean-variance equality of Poisson counts. Second, the assumption of a linear influence of covariates on the log-transformed expectation may be too rigid. In fact, for some applications the detection of changes in sensitivity with respect to a covariate is fundamental. 

In this work, we tackle a particular case of the first problem where overdispersion is (partly) due to an excess of zero counts. We assume a binary random effect which determines affiliation of observations  to a class of structural zeros or alternatively to the class of ordinary counts. Therefore, we use Zero Inflation (ZI) which proposes a mixture of a Dirac mass in $0$  with the respective count distribution, see \cite{lambert1992zero}. 
Switching from Poisson to a negative binomial distribution can take account of additional overdispersion in the ordinary component. Applications of ZI regression are numerous in the medical context, e.g. in public health studies like  \cite{boehning1999zero} and \cite{cheung2002zero}.  Other, more uncommon applications are e.g. software fault prediction in \cite{khoshgoftaar2001application} and patent outsourcing rates in \cite{czado2007zero}.  

To deal with the second problem of non-linearities, we resort to B-splines (cf. \cite{deBoor2001}). Spline-based approaches have become common for regression models in statistics, see the presentation \cite{eubank1988spline}   on spline smoothing and non-parametric regression, and  the Multivariate Adaptive Regression Spline technique (\textit{MARS}) used for normal residuals by \cite{Friedman}. 
Whereas \textit{smoothing spline} methods use the set of data points as spline knots and impose a smoothness  penalty, \textit{regression spline} methods tend to favor a small number of spline knots. For the latter methods, variable knots may be free knots of parametric nature or knots that are chosen by an adaptive method. The smoothness penalty introduced in the intermediate form of \textit{penalized regression splines} allows to use a higher number of knots  without overfitting. In summary, splines offer a semi-parametric setting such that the effective model dimension remains controllable 

Our approach replaces a linear regression coefficient by a spline curve determined by its degree of curvature, its knots and the coefficients of the resulting B-spline basis.  
A variant of the Evolutive Boundary Knots approach (\textit{EBOK}, \cite{Molinari}) is applied in order to use adaptive knots in the numerical calculation of the maximum likelihood estimator.

The introduced models for count data regression  unite interpretability of parameters and  high flexibility with respect to non-standard behavior, but avoid numerical complications by constraining knots. Moreover, iterating the EBOK method with different initial constraints leads asymptotically to globally optimal knots (see \cite{Molinari}). This class of models  can be classified among the large class of Generalized Additive Models (GAM; \cite{hastie1990generalized}).

In the remainder of the paper, we first present details of the modelisation in the following section. 
Issues and the algorithm related to the spline modeling approach are discussed in Section \ref{sec:opt}, followed by remarks on model selection criteria in Section \ref{sec:sel}.  The simulation study in Section \ref{sec:sim} sheds light on model choice and goodness-of-fit issues, with particular focus on the selection of a spline space. We apply the presented models to caries count data of $12$-year old French children in Section \ref{sec:appl}, with the Body-Mass-Index as potentially non-linear continuous covariate. Open issues and further possible developments are subject of Section \ref{sec:concl}.
\section{The Model}
\label{sec:model}
The Zero-Inflated Poisson Process $\zip (\mu, \pi)$  with parameters $\mu>0, \pi \in [0,1)$ has probability mass function (pmf)
\[
\left\{ \begin{array}{ll}
\pi + (1-\pi) \exp(-\mu) &  k=0 \\
(1-\pi)\frac{\mu^k}{k!}\exp(-\mu) &  k=1,2, ... 
\end{array} \right.
\]
The Zero-Inflated Negative Binomial Process $\zinb(\mu, \nu, \pi)$  with expectation $\mu>0$, dispersion parameter $\nu>0$ and   $\pi\in[0,1)$ has  pmf
\[
\left\{ \begin{array}{ll}
\pi + (1-\pi)\left(\frac{\nu}{\mu+\nu}\right)^{\nu} & k=0 \\
(1-\pi)\frac{\Gamma(k+\nu)}{k!\, \Gamma(\nu)} \frac{\nu^\nu\, \mu^k}{(\mu+\nu)^{k+\nu}}& k=1,2, ...
\end{array} \right.
\]
We denote univariate B-splines by $N_{i,d}(\,\cdot\,, \Delta), i = 1,...,m+d+1$, with a grid $\Delta$ of knots on a finite interval $[a,b]$, boundary knots $a,b$ and the degree $d$ of spline smoothness. The degree $d=1$ corresponds to  linear splines and $d=3$ to cubic splines.

We follow Lambert's model of two regression equations as presented in \cite{lambert1992zero}, allowing covariates to influence on the probability of a structural zero and on the count expectation. The model  for $n$ observations $Y_i$ is 

\[
\left\{ \begin{array}{ll}
& Y_i \mid (\mathbf{X_i},\mathbf{Z_i}) = (\mathbf{x_i},\mathbf{z_i}) \sim (\mu(\mathbf{x_i}), \pi(\mathbf{z_i}))\quad\mbox { independent}\\
& \log \mu(\mathbf{x_i}) = g^c(\mathbf{x_i}, \mathbf{\beta^c}) \\
& \mathrm{logit}\, \pi(\mathbf{z_i}) = g^z(\mathbf{z_i}, \mathbf{\beta^z}) 
\end{array}\right.
\]
The covariate vectors $\mathbf{x_i}$, $\mathbf{z_i}$ may coincide, e.g. if each component depends on all available covariates. Dependence on covariates can be modeled either by a linear regression coefficient  or by a spline function for continuous covariates. I.e.,  
\[
	\label{partieLin} 
	g^c(\mathbf{x_i}, \mathbf{\beta^c})= \sum\limits_{j=1}^{m^c}  g_j^c(x_{ij}, \beta_j^c), \quad g^z(\mathbf{z_i}, \mathbf{\beta^z})= \sum\limits_{j=1}^{m^z}  g_j^z(z_{ij},\beta_j^z)
\]
where  for $ind \in \{c, z \}$ either 
\[ 
g_j^{ind}(u, \beta_j)=u\,\beta_j 
\] 
or
 \[
 g_j^{ind}(u, \beta_j)= g_j^{ind}(u, \beta_j, \Delta_{j}) =\sum_l\beta_j^{(l)} N_{l,d}(u, \Delta_{j}).
 \]
 If splines are applied for $m>1$ covariates $x_{\cdot, j}, j\in J$, we remove $m-1$ of the terms $\beta_j^{(1)}N_{1,d}(u,  \Delta_{j})$, $j\in J$, to ensure an identifiable model: The "partition of $1$"-property $\sum N_{l,d}\equiv 1$ gives
 \[
 \sum_l\beta_j^{(l)} N_{l,d}(u, \Delta_{j}) = \beta_j^{(1)}+ \sum\limits_{l>1}(\beta_j^{(l)}- \beta_j^{(1)})N_{l,d}(u, \Delta_{j}), 
 \]
  so we combine the $m$ constants $\beta_j^{(1)}$ into one single coefficient  $\tilde{\beta}_{j_0}^{(1)}=\beta_{j_0}^{(1)}$, and for the other coefficients we replace $\beta_j^{(l)}$ by   $\tilde{\beta}_j^{(l)}=\beta_j^{(l)}- \beta_j^{(1)}$.  If we apply the linear modelisation instead of a spline curve in one of the components, as usual a regression constant can be added. 
  
For cubic splines,  we can optionally impose that second-order derivatives of a spline curve  are zero at the boundary knots $a,b$  of the respective covariate's domain -- resulting in a \textit{natural cubic spline}. This counteracts the strong near-boundary variability  of non-linear spline curves and reduces the number of free B-spline coefficients by $2$. The spline model is flexible in the sense that other algorithmically tractable constraints on the form of the spline curve could be imposed. 
  
The (semi-)parametric nature of the model with coefficients and knots results in models that are flexible and remain conveniently  interpretable and comparable, see also the remarks on model selection later on. Quantities like derivatives are analytically accessible. 
 
We fit this model by maximum-likelihood estimation of the $\beta$-coefficients, of knot locations $\Delta$ if considered as variable  and of  additional parameters that are constant for all observations, e.g. the extent of overdispersion $\nu$ of ZINB. Other links than the proposed ones may be chosen if convenient. The proposed model is of additive type,  but extensions with more refined multivariate spline techniques can deal with interaction effects of covariates. 

\section{Optimal Knot Locations}
\label{sec:opt}
In general, the set-up of fixed knots is an arbitrary restriction of the set of available spline curves. In particular, the B-spline basis may contain elements that merely improve  the fit and model dimension is blown up by insignificant parameters. For instance, think of  an equidistant grid of knots which performs poorly if there are large regions of sparse observations in the covariate domain.      
However, optimizing  knot locations makes numerical optimization more challenging for the following reasons:
\begin{itemize}
\item The number of free parameters in ``good'' models tends to increase.
\item Analytic partial derivates with respect to knots are difficult to establish; hence we use numerical derivatives. 
 \item Knots influence differently on the likelihood in comparison with $\beta$-coefficients.
 \item In particular, many suboptimal stationary points which correspond  to coinciding knots may arise (see \cite{Jupp} for this \textit{lethargy} property), rendering non-constrained knot optimization infeasible.
\end{itemize}

We use an Evolutive Bounded Optimal Knots algorithm (see \cite{Molinari}) that imposes iteratively adaptable interval constraints on variable knots, based on a partition of a covariate domain $[a,b]$. In general, the resulting optimized knots are not globally optimal, but iterating the procedure with different initial interval constraints would lead to convergence to the optimal set of knots.   To avoid coinciding knots and the associated discontinuities in spline functions, a minimal positive distance between two knots can be defined. 
In the following, we present and apply the algorithm for only one spline-modeled covariate; the extension to an additive model as introduced in Section \ref{sec:model} is straightforward.  However, for the inclusion of interaction terms in other    multivariate spline approaches, more complex adaptation steps become an issue. We further exclude iteration with different initial knot constraints and propose instead initial knots corresponding to equiprobable quantiles such that segments between knots contain the same number of observations.    
We shortly outline the functioning of the algorithm:
\begin{itemize}
 \item Initial knots are determined according to equiprobable quantiles of the covariate.
   \item Each knot can vary within the boundaries of an interval, also called its box. Boxes for different knots do not overlap, and usually we separate boxes by a small distance to avoid exactly coinciding knots. 
  Initial boundaries are fixed to the center of the segment between two initial knots. 
     \item We determine initial B-spline coefficients and other estimated parameters  by numerical maximisation of the likelihood  with the fixed initial knots. 
\item We maximize likelihood numerically with respect to box-constrained knots with the initial parameter estimates as starting point. 
\item If one of the knots coincides with a boundary of its box, we shift this boundary such that it is in the middle between this knot and its neighboring knot. The adjacent boundary of the box of the neighboring knot is shifted accordingly such that boxes do not overlap and the separating distance is preserved. 
\item If one of the knots coincides with a boundary, but the neighboring knot is too close to shift this boundary, we do nothing. Usually this means that there is a sharp break in the curvature of the fitted spline curve. 
 \item If box boundaries have been shifted, we iterate the procedure with the adapted box configuration.
\item If no boundaries are to be shifted, we stop the iteration, and the last set of knots and estimated parameters defines the fitted model. 
\end{itemize}

\section{Model selection}
\label{sec:sel}
We consider AIC and BIC  for model selection. Moreover, (cross-validation) residuals allow us to evaluate a model's predictive power. Difficulty lies in determining the dimension of spline models. With a parametric interpretation of spline coefficients and knots, we count each estimated scalar as a model parameter.  So for each spline curve included in the model, we add to the model dimension the sum $m+d+1 +m_f$ of the number $m$ of inner knots, of the spline order $d+1$ ($2$ for linear splines, $4$ for cubic splines) and of the number $m_f\leq m$ of free inner knots. This model dimension corresponds to $d+1$ initial parameters associated to the left endpoint $a$ of the covariate interval $[a,b]$, to $m$ change-points of which $m_f$ are free parameters, and  to the $m$ respective changes in the $(d-1)$th derivative. For instance,  a linear free-knot spline $s$ with $d=1$ possesses an initial level $s(a)$, an initial sensitivity $s'(a)$ and  $m=m_f$ breakpoints with the respective sensitivity changes. If natural cubic splines are applied, the model dimension is reduced by $2$. \\
Spline spaces include all linear functions, but the penalisation of higher model dimension would favor the simpler models with a linear coefficient when non-linear effects are not strong. If the sample is large, BIC penalizes strongly in the semi-parametric setting of splines which may include  a rather high number of parameters compared to purely parametric models. This seems disproportionate such that we prefer AIC, a choice which is substantiated through the simulation study in the following section. In  general, one should be cautious when comparing models with different nature of the respective parameters. In the literature, no consensus is yet found on how to assess the effective model dimension when splines are involved.  
Further model checks can be based on a more profound analysis of residuals.  

In the application to dental sanity data we opted for a preselection of the models performing best in terms of AIC among a wide variety of available models. To obtain our designated "winner model" and to elude selection bias, we checked these models' predictive power by minimizing a cross-validation mean residual error.

\section{A simulation study}
\label{sec:sim}
We focus on two issues:
\begin{itemize}
	\item Study (1): Retrieve the B-spline coefficients of a simulated model when the spline space is known.
	\item Study (2): Find the best-fitting spline model when the simulated model is not equivalent to a spline model.
\end{itemize}
In each case, we use ZIP regression. We assess the (relative) goodness-of-fit for  different models based on AIC, BIC and cross-validation mean residual error  with respect to raw residuals (MRE). We randomly choose $20$ non-overlapping subsamples of equal size for cross-validation to avoid the costly leave-one-out procedure. Moreover, we apply the supremum- and $L^1$-norms to measure the deviation of fitted spline curves from the simulated ones.

\begin{figure}
  \centering
  \includegraphics[width=3.5in]{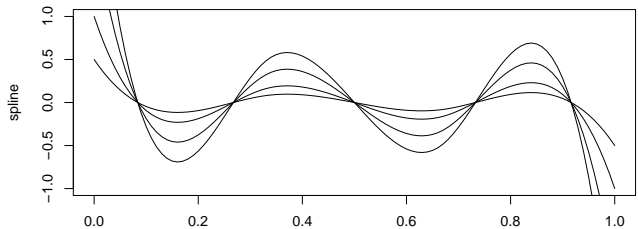}
\caption{Spline curves for simulated models of Study $(1)$.}
\label{study1}
\end{figure}

In Study (1), we restrict the analysis to the more intricate cubic splines which usually lead to stronger variability in fitted curves than linear splines, particularly near the boundaries. We generate $100$ simulations of the model
\begin{equation}
\label{zipSim}
	 X_i \sim U[0,1], \ 
	 Y_i | X_i=x_i \sim \mathrm{ZIP}\left(\log^{-1}f_{c}(x_i), \mathrm{logit}^{-1} f_z(x_i)  \right),
\end{equation}
with $i=1,...,200$, where $f_c$ is a cubic spline  
\[
f_c(x)=\sum\limits_{i=1}^6 \alpha \beta_i N_{i,k}(x)
\]
with marginal knots $0$ and $1$ and with $2$ equidistant interior knots $1/3$ and $2/3$, and 

\[
f_z(x)=\beta_0^z+\beta_1^zx=1-x\ .
\]
The coefficients $\alpha \beta_i$ are chosen such that
\[ 
\beta_i = -\left(-1\right)^{i} \quad (i=1, ..., 6)\ ,  
\]
and $\alpha\in \{0.5, 1, 2, 3 \}$ determines how strongly the spline function oscillates, see Figure \ref{study1}. Tables~\ref{t:study1.1} to~\ref{t:study1.4} summarize for the four given  values of $\alpha$  the goodness-of-fit statistics for linear fits and cubic spline fits with up to three equistant fixed knots in the count component. We show median values and standard deviations over the $100$ simulations for the proposed criteria (note that the standard deviation coincides for AIC and BIC). Additionally, for AIC, BIC and MRE, the number of times a model performed best is given. Median values and standard deviations for the estimated regression constants and coefficients of the zero component are also included. 
 
The case $\alpha=0.5$ with the weakest oscillation is best fitted with a linear model. Even the norm criteria tend to favor the linear model. In general, curves that are "close" to linear are better fitted by linear models, since higher model dimension and increased variability make the spine fit less appealing. With $\alpha$ increasing, the actual model is better reconstructed.  

AIC, MRE and the $L^1$-norm show similar preferences, whereas BIC has a stronger preference for the simpler linear model. The fitted linear coefficients of the zero component show some bias in some cases, but no other anomalies. 

In Study (2), we keep the model from Study (1) with $50$ simulations,  where we change $f_c$ to the non-spline function
\[
f_c(x)=1+\sqrt{x} + \sin(4\pi x),\ x \in [0,1]\ , 
\]
which describes a periodic phenomenon  $\sin(4\pi x)$  with a non-linear trend $1+\sqrt{x}$.

We fit linear splines and cubic splines with fixed and adaptive knots (EBOK method).  Tables~\ref{t:study2.1} to ~\ref{t:study2.4} show the results. 
MRE preference is scattered over the range of models such that there is no clear preference. We always find a good correspondance between AIC with $L^1$-norm and partly with the supremum norm, so AIC serves well in finding the best model. BIC tends to simpler models.  Variable knots tend to improve the examined criteria, and less knots are needed to obtain the optimal value of a criterion. However, an improvement in best attainable BIC values seems questionable when switching to variable knots.

We conclude that, in practice,  the presented class of ZI spline regression models allows for fitting to data with the proposed fitting algorithm which is sufficiently robust. BIC works well if non-linearities are strong, in other cases we advise AIC, which can be complemented by the analysis of residuals, in particular of MRE.

\section{Application to dental sanity data}
\label{sec:appl}
\begin{figure}
  \centering
  \begin{tabular}{cc}
   \includegraphics[width=3in]{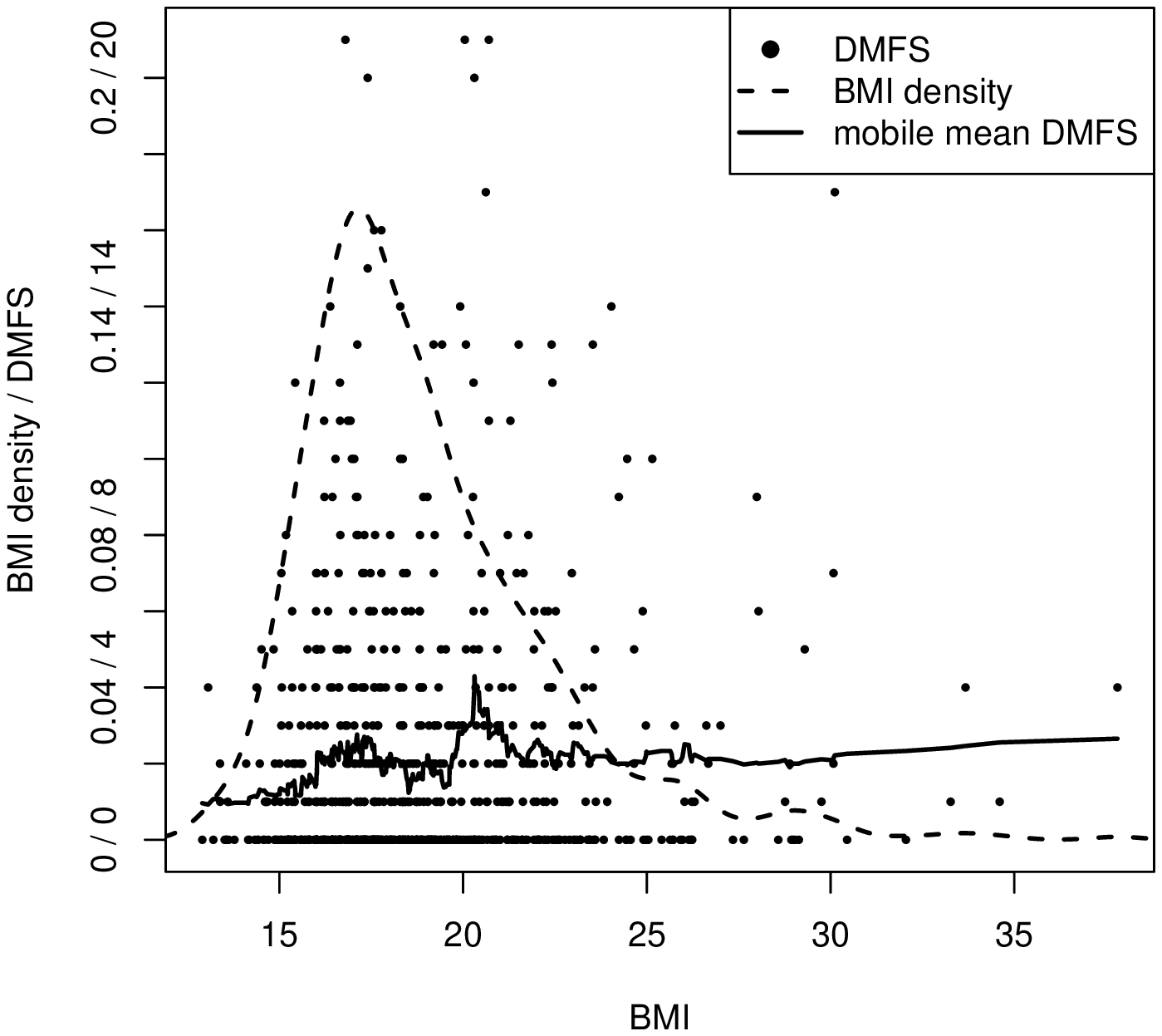}
& 
   \includegraphics[width=1.5in]{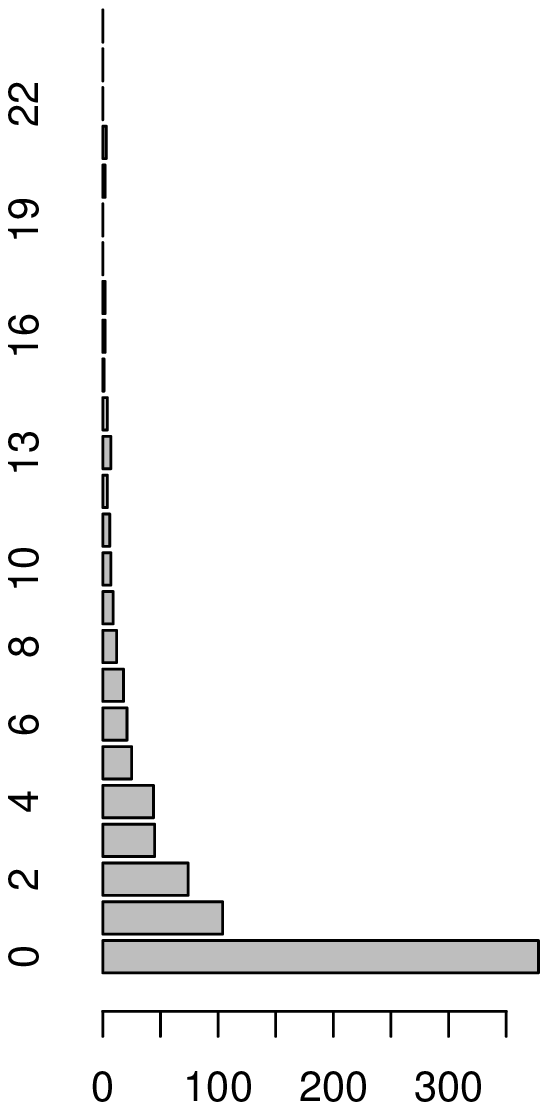}
  \end{tabular}
\caption{Left: Scatterplot of DMFS and BMI data. Right: DMFS frequencies.  }
\label{dataDesc}
\end{figure}

Finding appropriate caries prediction models is a recurring subject in odontology, see the review of \cite{powell2007} and the meta-analysis of \cite{harris2004} concerning risk factors for dental caries in young children. A commonly accepted quantification of caries incidence for an individual is the DMFS index, i.e. the count of decayed, missing and  filled teeth surfaces. 
We have at our disposal a data set for $768$  French children aged $12$ years. We use the variables DMFS index, weight, height, school type (public vs. private), sugar consumption (low vs. high) and consumption of sweetened drinks (low vs. high). 
An important quantity with respect to an individual's health is the so-called \textit{Body Mass Index} BMI. It is obtained by dividing the weight (in kg) by the squared height (in m), thus defining a continuous scale ranging from underweight to obesity (cf.  \cite{Ferrara}). 
A discussion of the association between caries counts and BMI  in a more traditional statistical context is given in \cite{Tramini}. There the question was raised if non-linear regression techniques could better identify a possible dependence on BMI, since no definite conclusion on significance of BMI could be drawn. Figure~\ref{dataDesc} shows a scatterplot of DMFS and BMI data, including a kernel density estimate of BMI and a mobile mean of DMFS with respect to BMI. The mobile mean is calculated over a neighborhood of $25$ BMI values at each side. Naively, one might expect either no effect or e.g. a monotonic increase of mean DMFS values with increasing BMI. Yet the mobile mean shows two peaks and a dent in-between in a region of high density of BMI observations, with an overall slightly increasing trend. 
Moreover, zero observations are abundant in DMFS values. The ZI spline regression models are therefore appropriate candidates for a formal modelisation of these data.
We fitted a wide variety of models determined by all combinations of the following options: 
\begin{itemize}
	\item with vs. without ZI
	\item BMI: linear in the count or the structural zero component or in both
	\begin{itemize}
		\item regression constant only vs. regression constant and  coefficient
	\end{itemize}
	\item with vs. without a coefficient for school type
	\item with vs. without a coefficient for sugar consumption
	\item with vs. without a coefficient for consumption of sweetened drinks
	\item spline regression
	\begin{itemize}
		\item adaptive knots vs. fixed equiprobable quantile knots
		\item linear vs. cubic vs. natural cubic splines
	\end{itemize}
\end{itemize}
For spline curves, we considered at least one knot and a maximum number of knots such that the optimal AIC values could be identified. Since we expect non-linear effects to be rather weak, we neglect BIC in accordance with the results of the preceding simulation study.
We restrain the following analysis to finding a single model that fits well. Depending on what we want to know about the process that generates the data, subsets of the considered model classes may be of interest -- e.g. linear variable knot models to identify breakpoints. 
We summarize some noticeable caracteristics of the fitted models:
\begin{description}
\item Negative binomial models are clearly superior to Poisson models. So even after including ZI into the model, there remains overdispersion in the count component. 
\item A linear spline model with ZI and $3$ fixed interior knots takes the top position with respect to AIC. 
\item The factor covariates turn up in the top AIC ranks. In particular, the consumption of sweetened drinks is included in the two top models. 
\item We mention that the strong penalization of BIC favors simple linear models without ZI. In fact, the nonconditional model with no influence of the covariates turns out best.
\end{description}

\begin{figure}
  \centering
  \begin{tabular}{cc}
   \includegraphics[width=2.25in]{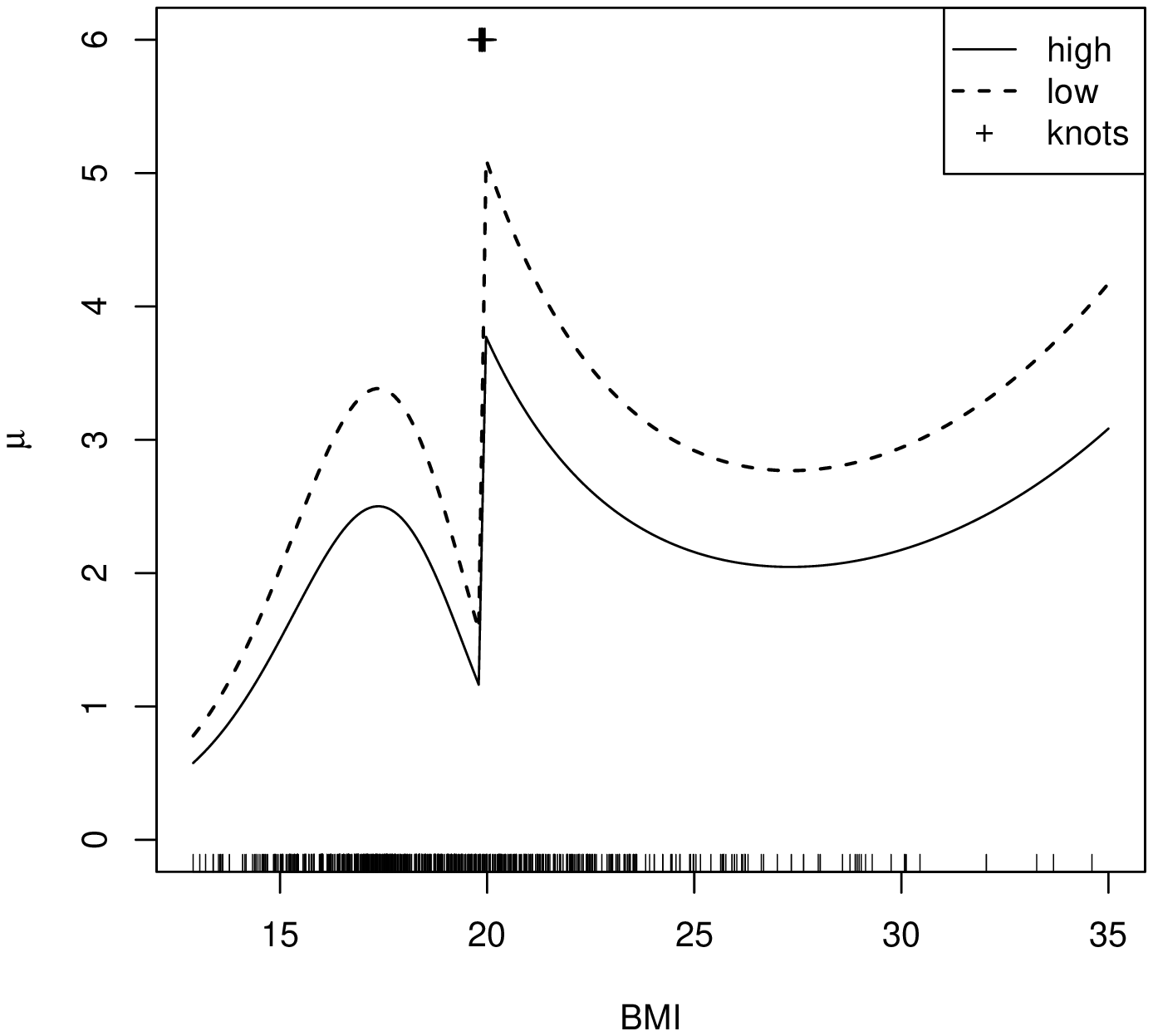}
  &
   \includegraphics[width=2.25in]{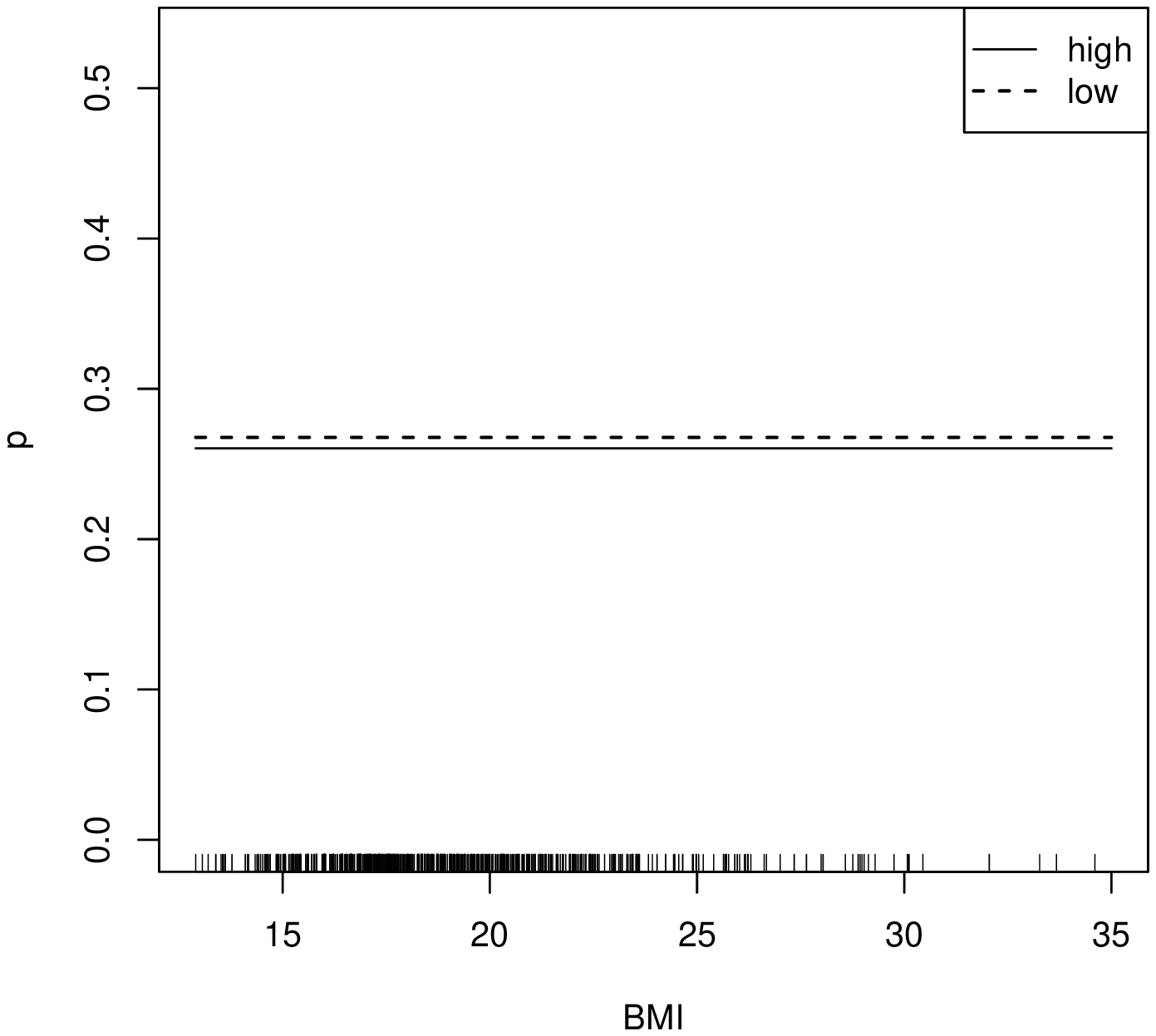}
  \end{tabular}
    \caption{Winner model. Includes the factor for the consumption of sweetened drinks and $4$ (almost coinciding) knots. The rug indicates the observed DMFS data.}
\label{bestfreeAIC}
\end{figure}

\begin{figure}
\centering
  \includegraphics[width=3in]{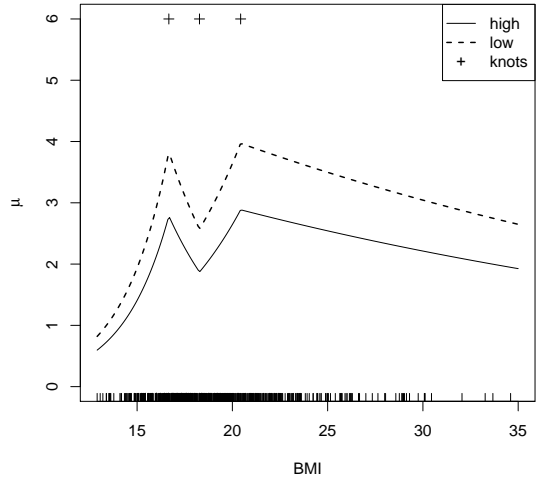}
  \caption{Best AIC model. With factor for consumption of sweetened drinks and $3$ (fixed) knots. The rug indicates the observed DMFS data.}
\label{bestAIC}
\end{figure}

We found that mean residual errors with leave-one-out cross-validation (MRE) for the top $20$ AIC models vary little between $2.410$ and $2.433$. We look closer at the top AIC model and the best model with respect to MRE  which we choose as "winner",  see Figures~\ref{bestfreeAIC}  and ~\ref{bestAIC} for predictions of structural zero probability and mean count $\mu$. The zero probabilities do not depend on BMI. 
In both models, the factor for high consumption of sweetened drinks is included and  increases the count expectation clearly. Curiously, the probability of structural zeros increases slightly with high consumption. We conjecture that actually there is no effect of this factor on structural zeros, i.e., that the corresponding coefficient is $0$. 
The winner model is the only variable knot model among the $20$ preselected models. The plot for the structural zero probability is omitted since it is virtually identical to the one in Figure \ref{bestAIC}. The model applies  natural splines and $4$ knots that almost coincide, with $\varepsilon=0.001\times\mathrm{range} (\mathrm{BMI})$ as minimal knot distance in the EBOK algorithm. There is an abrupt jump in the predicted values around the interior knot locations. 
In summary, both models reproduce the two peaks and a dent in-between, as revealed by the application of a mobile mean. The winner model's abrupt jump in the dent is even more baffling. An explanation for this interesting finding has yet to be established by odontological researchers.

\section{Conclusion and outlook}
\label{sec:concl}
We presented a class of models that provides high flexibility for regression analysis of count data. In particular, we allow for a binary effect that decides on membership to structural zero observations. The spline approach serves to identify non-linear relations. This is of great importance in fields like medicine or disease control in order to correctly assess the influence of risk factors.
When computational complexity of variable knots is manageable, the goodness  of the spline fit tends to improve even with less knots than in the fixed knot scenario.   
A refinement of model choice may prevent a biased selection of the winner model which is due to the usually large number of eligible models. 
In general, multiple tests and the comparison of different goodness-of-fit criteria help improve the model selection procedure. Criteria for high-dimensional model spaces like the slope heuristic could be considered. 
Since any parametric linear model is equivalent to a semi-parametric spline model, the decision if a spline model fits better could be based on procedures operating only on spline model fits. Then the pooling of parametric and semi-parametric models, as done in the presented application, can be avoided for model selection. 
We showed the utility of such model fitting in an application to dental sanity data where classical GLM models and ZI-GLM models failed to reveal a significant influence of the Body-Mass-Index on caries counts.  

\bibliographystyle{plain}
\bibliography{ref}

\clearpage
\begin{table}
  \caption{Study $(1)$: Linear fit and cubic spline fits for $100$ simulations.  
  }
\label{t:study1.1}
\centering
\begin{tabular}{ll|llll}
\toprule
 \multicolumn{2}{c|}{$\alpha=0.5$  } & lin.& $1$& $2$& $3$\\
 \midrule
$\|\cdot \|_\infty$ & best & $\mathbf{37}$& $25$& $16$& $22$\\
 & med. & $0.4787$& $0.4796$& $\mathbf{0.4633}$& $0.4653$\\
 & sd & ($0.2536$)& ($0.3781$)& ($0.4012$)& ($0.4441$)\\
$\|\cdot \|_1$ & best & $\mathbf{89}$& $5$& $5$& $1$\\
 & med. & $\mathbf{0.1928}$& $0.2971$& $0.3317$& $0.364$\\
 & sd & ($0.1147$)& ($0.1243$)& ($0.126$)& ($0.1495$)\\
 \hline 
MRE & best & $\mathbf{35}$& $28$& $21$& $16$\\
 & med. & $0.5871$& $0.5801$& $0.5809$& $\mathbf{0.5787}$\\
 & sd & ($0.07391$)& ($0.07362$)& ($0.07337$)& ($0.07315$)\\
 \hline 
 AIC & best & $\mathbf{71}$& $16$& $6$& $7$\\
 & med. & $\mathbf{330.3}$& $330.5$& $331.9$& $331.3$\\
 & sd & ($32.16$)& ($31.72$)& ($31.83$)& ($31.93$)\\
 BIC & best & $\mathbf{100}$& $0$& $0$& $0$\\
 & med. & $\mathbf{343.5}$& $353.6$& $358.3$& $361$\\
 \hline 
$\beta_0^{z}$ & med. & $0.9773$& $0.8133$& $0.7484$& $0.7664$\\
 & sd & ($0.6067$)& ($0.6447$)& ($0.6587$)& ($0.6576$)\\
$\beta_1^{z}$ & med. & $-0.8938$& $-0.7376$& $-0.7058$& $-0.7598$\\
 & sd & ($1.098$)& ($1.667$)& ($1.688$)& ($1.658$)\\
 \bottomrule
\end{tabular}
\end{table}
\begin{table}
  \caption{Study $(1)$: Linear fit and cubic spline fits for $100$ simulations.  
  }
\label{t:study1.2}
\centering
\begin{tabular}{ccllll}
\toprule
 \multicolumn{2}{c|}{$\alpha=1$ }   & lin.& $1$& $2$& $3$\\
\midrule
$\|\cdot \|_\infty$ & best 
& $12$& $15$& $\mathbf{48}$& $25$\\
 & med.  
& $1.001$& $0.8161$& $\mathbf{0.5528}$& $0.6432$\\
 & sd  
& ($0.3142$)& ($0.451$)& ($0.4037$)& ($0.6481$)\\
$\|\cdot \|_1$ & best  
&$\mathbf{64}$& $9$& $20$& $7$\\
 & med. 
 & $\mathbf{0.2542}$& $0.3256$& $0.3409$& $0.3907$\\
 & sd  
 &($0.1047$)& ($0.1157$)& ($0.1447$)& ($0.1531$)\\
 \hline 
MRE & best 
 &$25$& $20$& $\mathbf{35}$& $20$\\
 & med. 
 & $0.5982$& $0.5989$& $\mathbf{0.5921}$& $0.5956$\\
 & sd 
 & ($0.06855$)& ($0.06817$)& ($0.06734$)& ($0.06885$)\\
 \hline 
 AIC & best 
 & $\mathbf{46}$& $12$& $28$& $14$\\
 & med.  
 &$329.7$& $329.6$& $\mathbf{328.1}$& $329.3$\\
 & sd 
 &($29.71$)& ($29.41$)& ($29.32$)& ($28.99$)\\
 BIC & best 
  &$\mathbf{98}$& $0$& $2$& $0$\\
 & med.  
& $\mathbf{342.9}$& $352.7$& $354.5$& $358.9$\\
 \hline 
$\beta_0^{z}$ & med. 
 &$0.9504$& $0.8906$& $0.771$& $0.7398$\\
 & sd 
&  ($0.6368$)& ($0.5333$)& ($0.5237$)& ($0.5268$)\\
$\beta_1^{z}$ & med. 
& $-0.7811$& $-0.6593$& $-0.6756$& $-0.6786$\\
 & sd 
& ($1.279$)& ($1.092$)& ($1.079$)& ($1.096$)\\
 \bottomrule
\end{tabular} 
\end{table}

\begin{table}
  \caption{Study $(1)$: Linear fit and cubic spline fits for $100$ simulations.  
  }
\label{t:study1.3}
\centering
\begin{tabular}{ll|llll}
\toprule
 \multicolumn{2}{c|}{$\alpha=2$ } & lin.& $1$& $2$& $3$\\
 \midrule
\hline$\|\cdot \|_\infty$ & best & $2$& $2$& $\mathbf{59}$& $37$\\
 & med. & $1.721$& $1.454$& $\mathbf{0.5498}$& $0.6712$\\
 & sd & ($0.4325$)& ($0.7486$)& ($0.4782$)& ($0.6693$)\\
$\|\cdot \|_1$ & best & $16$& $6$& $\mathbf{61}$& $17$\\
 & med. & $0.4552$& $0.4504$& $\mathbf{0.347}$& $0.3867$\\
 & sd & ($0.08129$)& ($0.1226$)& ($0.1642$)& ($0.1893$)\\
 \hline 
MRE & best & $22$& $7$& $\mathbf{38}$& $33$\\
 & med. & $0.6035$& $0.6115$& $\mathbf{0.5969}$& $0.6$\\
 & sd & ($0.08173$)& ($0.07994$)& ($0.07858$)& ($127.2$)\\
 \hline 
 AIC & best & $12$& $6$& $\mathbf{59}$& $23$\\
 & med. & $323.3$& $322.1$& $316.6$& $\mathbf{316.6}$\\
 & sd & ($32.37$)& ($30.98$)& ($29.35$)& ($29.68$)\\
 BIC & best & $\mathbf{73}$& $2$& $25$& $0$\\
 & med. & $\mathbf{336.5}$& $345.2$& $343$& $346.3$\\
 \hline 
$\beta_0^{z}$ & med. & $1.255$& $1.08$& $0.9983$& $0.972$\\
 & sd & ($0.5357$)& ($1.001$)& ($0.523$)& ($0.5195$)\\
$\beta_1^{z}$ & med. & $-1.269$& $-0.952$& $-1.023$& $-1.057$\\
 & sd & ($1.122$)& ($1.46$)& ($1.126$)& ($1.108$)\\
 \bottomrule
\end{tabular}
\end{table}

\begin{table}
  \caption{Study $(1)$: Linear fit and cubic spline fits for $100$ simulations.  
  }
\label{t:study1.4}
\centering
\begin{tabular}{ll|llll}
\toprule
 \multicolumn{2}{c|}{$\alpha=3$} &
 lin.& $1$& $2$& $3$\\
 \midrule
 $\|\cdot \|_\infty$ & best &
 $0$& $0$& $\mathbf{63}$& $37$\\
 & med. &
  $2.525$& $2.569$& $0.5505$& $\mathbf{0.5438}$\\
 & sd &
  ($0.4793$)& ($1.237$)& ($0.4578$)& ($0.3987$)\\
$\|\cdot \|_1$ & best & 
$2$& $1$& $\mathbf{85}$& $12$\\
 & med. & 
 $0.7175$& $0.6292$& $\mathbf{0.2804}$& $0.3449$\\
 & sd & 
 ($0.1066$)& ($0.1043$)& ($0.162$)& ($0.1786$)\\
 \hline 
MRE & best &
 $20$& $4$& $\mathbf{39}$& $37$\\
 & med. &
  $0.7665$& $0.7564$& $0.7341$& $\mathbf{0.7323}$\\
 & sd & 
 ($0.1283$)& ($0.1868$)& ($0.1112$)& ($0.1171$)\\
 \hline 
 AIC & best &
  $2$& $1$& $\mathbf{71}$& $26$\\
 & med. & 
 $363.3$& $352.9$& $\mathbf{335.8}$& $338.6$\\
 & sd &
  ($42.61$)& ($35.84$)& ($31.55$)& ($31.56$)\\
 BIC & best &
  $19$& $1$& $\mathbf{77}$& $3$\\
 & med. & 
 $376.5$& $376$& $\mathbf{362.2}$& $368.3$\\
 \hline 
$\beta_0^{z}$ & med. & 
$1.519$& $1.083$& $0.9298$& $0.8992$\\
 & sd &
  ($0.6304$)& ($0.5472$)& ($0.505$)& ($0.4898$)\\
$\beta_1^{z}$ & med. &
 $-1.662$& $-0.8224$& $-1.091$& $-1.034$\\
 & sd &
  ($2.16$)& ($1.358$)& ($1.017$)& ($0.9697$)\\
 \bottomrule
\end{tabular}
\end{table}

\begin{table}
 \centering
  \caption{Study $(2)$: Linear fit and linear spline fits for $50$ simulations.}
\label{t:study2.1}
\centering
\begin{tabular}{ll|lllllll}
\toprule
 \multicolumn{2}{c|}{{\bf fixed knots}} & lin.& $1$& $2$& $3$& $4$& $5$& $6$\\
 \midrule
$\|\cdot \|_\infty$ & best & $0$& $0$& $0$& $0$& $2$& $12$& $\mathbf{36}$\\
 & med. & $1.321$& $1.429$& $1.303$& $1.254$& $0.7911$& $0.6064$& $\mathbf{0.4843}$\\
 & sd & ($0.08845$)& ($0.1048$)& ($0.2739$)& ($0.3453$)& ($0.3474$)& ($0.2732$)& ($0.2471$)\\
$\|\cdot \|_1$ & best & $0$& $0$& $0$& $0$& $1$& $13$& $\mathbf{36}$\\
 & med. & $0.5916$& $0.588$& $0.4133$& $0.5209$& $0.2603$& $0.1994$& $\mathbf{0.1736}$\\
 & sd & ($0.02193$)& ($0.02282$)& ($0.03472$)& ($0.03662$)& ($0.03817$)& ($0.0459$)& ($0.03317$)\\
 \hline 
MRE & best & $5$& $1$& $6$& $1$& $10$& $\mathbf{14}$& $13$\\
 & med. & $3.23$& $3.225$& $3.099$& $3.171$& $3.104$& $3.083$& $\mathbf{3.083}$\\
 & sd & ($0.3565$)& ($0.3487$)& ($0.2872$)& ($0.3025$)& ($0.2665$)& ($0.2525$)& ($0.2414$)\\
 \hline 
 AIC & best & $0$& $0$& $0$& $0$& $4$& $16$& $\mathbf{30}$\\
 & med. & $760.5$& $756.9$& $660.8$& $697.6$& $616.2$& $\mathbf{600.7}$& $602.2$\\
 & sd & ($69.02$)& ($66.07$)& ($52.34$)& ($49.92$)& ($45.74$)& ($40.39$)& ($39.84$)\\
 BIC & best & $0$& $0$& $0$& $0$& $11$& $\mathbf{20}$& $19$\\
 & med. & $773.6$& $773.4$& $680.6$& $720.7$& $642.6$& $\mathbf{630.4}$& $635.1$\\
 \bottomrule
\end{tabular}
\end{table}

\begin{table}
  \caption{Study $(2)$: Linear fit and linear spline fits for $50$ simulations.}
\label{t:study2.2}
\centering
\begin{tabular}{ll|llllll}
\toprule
 \multicolumn{2}{c|}{{\bf var. knots}} & $1$& $2$& $3$& $4$& $5$& $6$\\
 \midrule
 $\|\cdot \|_\infty$ & best & $0$& $0$& $1$& $16$& $15$& $\mathbf{18}$\\
 & med. & $1.387$& $1.286$& $1.14$& $0.4642$& $\mathbf{0.442}$& $0.4773$\\
 & sd & ($0.08547$)& ($0.3015$)& ($0.3556$)& ($0.2394$)& ($0.2146$)& ($0.2337$)\\
$\|\cdot \|_1$ & best & $0$& $0$& $0$& $\mathbf{24}$& $15$& $11$\\
 & med. & $0.5387$& $0.3356$& $0.2626$& $\mathbf{0.163}$& $0.1871$& $0.1849$\\
 & sd & ($0.01319$)& ($0.02739$)& ($0.06051$)& ($0.04077$)& ($0.04256$)& ($0.05947$)\\
 \hline 
MRE & best & $10$& $6$& $\mathbf{12}$& $9$& $3$& $10$\\
 & med. & $3.189$& $3.104$& $3.12$& $3.118$& $3.099$& $\mathbf{3.086}$\\
 & sd & ($0.3405$)& ($0.2687$)& ($0.248$)& ($0.2352$)& ($0.2144$)& ($0.2611$)\\
 \hline 
 AIC & best & $1$& $3$& $5$& $\mathbf{18}$& $14$& $9$\\
 & med. & $703.6$& $639.7$& $621.5$& $\mathbf{593}$& $593.5$& $598.3$\\
 & sd & ($57.8$)& ($45.73$)& ($43$)& ($39.36$)& ($35.93$)& ($39.84$)\\
 BIC & best & $2$& $3$& $8$& $\mathbf{21}$& $12$& $4$\\
 & med. & $723.4$& $666.1$& $654.5$& $\mathbf{632.6}$& $639.6$& $651$\\
 \bottomrule
\end{tabular}
\end{table}

\clearpage

\begin{table}
  \caption{Study $(2)$: Cubic spline fits for $50$ simulations. }
\label{t:study2.3}
 \centering
\begin{tabular}{ll|llllll}
\toprule
 \multicolumn{2}{c|}{{\bf fixed knots}} & $1$& $2$& $3$& $4$& $5$& $6$\\
 \midrule 
$\|\cdot \|_\infty$ & best & $0$& $\mathbf{15}$& $1$& $14$& $11$& $9$\\
 & med. & $1.309$& $0.3749$& $0.5663$& $\mathbf{0.3207}$& $0.3449$& $0.3494$\\
 & sd & ($0.7008$)& ($0.2088$)& ($0.35$)& ($0.2479$)& ($0.3341$)& ($0.3235$)\\
$\|\cdot \|_1$ & best & $0$& $11$& $0$& $14$& $\mathbf{15}$& $10$\\
 & med. & $0.5636$& $0.1609$& $0.2504$& $\mathbf{0.1347}$& $0.1353$& $0.1446$\\
 & sd & ($0.02085$)& ($0.03708$)& ($0.0466$)& ($0.04189$)& ($0.0518$)& ($0.05427$)\\
 \hline 
MRE & best & $8$& $\mathbf{14}$& $6$& $7$& $9$& $6$\\
 & med. & $3.238$& $3.103$& $3.154$& $3.101$& $\mathbf{3.088}$& $3.108$\\
 & sd & ($0.3993$)& ($0.2511$)& ($0.2568$)& ($0.2449$)& ($0.2374$)& ($0.2364$)\\
 \hline 
 AIC & best & $0$& $12$& $0$& $\mathbf{23}$& $11$& $4$\\
 & med. & $714.2$& $598.5$& $616.2$& $595.9$& $\mathbf{595.9}$& $599.2$\\
 & sd & ($57.14$)& ($42.12$)& ($39.69$)& ($39.18$)& ($38.6$)& ($38.8$)\\
 BIC & best & $0$& $\mathbf{37}$& $1$& $11$& $1$& $0$\\
 & med. & $737.3$& $\mathbf{624.8}$& $645.9$& $628.9$& $632.2$& $638.8$\\
 \bottomrule
\end{tabular}
\end{table}

\begin{table}
  \caption{Study $(2)$: Cubic spline fits for $50$ simulations. }
\label{t:study2.4}
 \centering
\begin{tabular}{ll|llllll}
\toprule
 \multicolumn{2}{c|}{{\bf var. knots} } & $1$& $2$& $3$& $4$& $5$& $6$\\
 \midrule 
$\|\cdot \|_\infty$ & best & $0$& $\mathbf{29}$& $8$& $4$& $4$& $5$\\
 & med. & $1.357$& $\mathbf{0.2841}$& $0.348$& $0.4517$& $0.4851$& $0.6122$\\
 & sd & ($0.7506$)& ($0.1916$)& ($1.514$)& ($1.926$)& ($2.918$)& ($1.886$)\\
$\|\cdot \|_1$ & best & $0$& $\mathbf{32}$& $10$& $3$& $4$& $1$\\
 & med. & $0.4675$& $\mathbf{0.126}$& $0.1382$& $0.1713$& $0.1883$& $0.2139$\\
 & sd & ($0.02244$)& ($0.04604$)& ($0.4086$)& ($0.2071$)& ($0.13$)& ($0.5814$)\\
 \hline 
MRE & best & $6$& $\mathbf{14}$& $5$& $7$& $7$& $11$\\
 & med. & $3.202$& $\mathbf{3.107}$& $3.112$& $3.135$& $3.126$& $3.121$\\
 & sd & ($0.3382$)& ($0.2447$)& ($0.4305$)& ($2.431$)& ($28.84$)& ($2285$)\\
 \hline 
 AIC & best & $0$& $\mathbf{31}$& $7$& $3$& $1$& $8$\\
 & med. & $692$& $\mathbf{598.6}$& $599.9$& $601.8$& $602.1$& $601.3$\\
 & sd & ($50.82$)& ($39.37$)& ($39.02$)& ($36.1$)& ($39.31$)& ($39.43$)\\
 BIC & best & $0$& $\mathbf{41}$& $1$& $1$& $0$& $7$\\
 & med. & $718.4$& $\mathbf{631.6}$& $639.5$& $647.9$& $654.9$& $660.7$\\
 \bottomrule
\end{tabular}
\end{table}
\end{document}